\documentclass{article} 
\usepackage{iclr2026_conference,times}

\usepackage{amsmath,amsfonts,bm}

\def\eqref#1{equation~\ref{#1}}

\def\1{\bm{1}}

\DeclareMathAlphabet{\mathsfit}{\encodingdefault}{\sfdefault}{m}{sl}
\SetMathAlphabet{\mathsfit}{bold}{\encodingdefault}{\sfdefault}{bx}{n}

\usepackage{hyperref}
\usepackage{url}
\usepackage{capt-of}
\usepackage[utf8]{inputenc} 
\usepackage[T1]{fontenc}    
\usepackage{url}            
\usepackage{booktabs}       
\usepackage{amsfonts}       
\usepackage{nicefrac}       
\usepackage{microtype}      
\usepackage{amsmath}
\usepackage{multirow}
\usepackage{subcaption}
\usepackage{pifont}
\usepackage{algorithm}
\usepackage{bbm}

\usepackage{textcomp}
\usepackage{stfloats}
\usepackage{mathabx}
\usepackage{amsthm}
\usepackage{enumitem}
\setlist{leftmargin=5mm}
\usepackage{wrapfig}
\usepackage{makecell}
\usepackage{wrapfig}
\usepackage{lipsum}
\usepackage{amsmath, amssymb, graphicx, xcolor, xspace}
\usepackage{booktabs}
\usepackage{algpseudocode}
\usepackage{bbm}
\usepackage{tcolorbox}  
\usepackage{float}
\tcbuselibrary{breakable} 
\usepackage[utf8]{inputenc}
\usepackage{tcolorbox}
\usepackage{microtype}
\newtcolorbox{mybox}[1][]{
    title=#1,
    fonttitle=\small,
    fontupper=\small,
    left=1mm,
    right=1mm,
    top=1mm,
    bottom=0mm,
}
\definecolor{darkblue}{rgb}{0, 0, 0.5}
\definecolor{darkred}{rgb}{0.72, 0.22, 0.27}
\definecolor{lightblue}{RGB}{129, 209, 241}
\definecolor{forestgreen}{RGB}{34, 139, 34}
\hypersetup{
    colorlinks=true,
    citecolor=darkblue,
    linkcolor=darkblue,
    urlcolor=darkblue,
}
\newcommand{\cmark}{\ding{51}}
\newcommand{\xmark}{\ding{55}}
\usepackage{inconsolata}

\usepackage{graphicx}
\usepackage[utf8]{inputenc} 
\usepackage[T1]{fontenc}    
\usepackage{url}            
\usepackage{booktabs}       
\usepackage{amsfonts}       
\usepackage{nicefrac}       
\usepackage{microtype}      
\usepackage{xcolor}         
\usepackage{amssymb}
\usepackage{amsmath}
\usepackage{multirow}
\usepackage{subcaption}
\usepackage{pifont}
\usepackage{amssymb}
\usepackage{algorithm}
\usepackage{bbm}
\usepackage{stfloats}
\usepackage{mathabx}
\usepackage{amsthm}
\usepackage[table]{xcolor}
\usepackage{enumitem}
\setlist{leftmargin=5mm}
\usepackage{graphicx}
\usepackage{wrapfig}
\usepackage{makecell}
\usepackage[dvipsnames]{xcolor}
\usepackage{wrapfig}
\usepackage{lipsum}
\usepackage{amsmath, amssymb, graphicx, xcolor, xspace}
\usepackage{booktabs}
\usepackage{algpseudocode}
\usepackage{bbm}
\newcommand{\up}[1]{\textcolor{green!55!black}{$\uparrow$ #1}}

\definecolor{gMMAU}{HTML}{C0D9ED}
\definecolor{gMMAR}{HTML}{F9E8C5}
\definecolor{gMMSU}{HTML}{CFBAD9}
\definecolor{gOVR}{HTML}{CFBAD9}
\usepackage{tikz}
\makeatletter
\newcommand{\DrawLine}{%
  \begin{tikzpicture}
  \path[use as bounding box] (0,0) -- (\linewidth,0);
  \draw[color=black,dashed,dash phase=2pt]
        (0-\kvtcb@leftlower-\kvtcb@boxsep,0)--
        (\linewidth+\kvtcb@rightlower+\kvtcb@boxsep,0);
  \end{tikzpicture}%
  }

\definecolor{violet}{RGB}{138, 43, 226}

\makeatother

\definecolor{citepcol}{HTML}{2DDC0E}
\definecolor{tableofcontent}{HTML}{E63E15}
\definecolor{urlcol}{HTML}{2470D8}
\definecolor{myorange}{RGB}{2, 142, 2}
\hypersetup{
    colorlinks=true,
    linkcolor=red,
    filecolor=magenta,      
    urlcolor=cyan,
}

\definecolor{barGray}{HTML}{D9D9D9} 
\NewDocumentCommand{\shibo}{ mO{} }
{\textcolor{pink}{\textsuperscript{\textit{Shibo}}\textsf{\textbf{\small[#1]}}}}
\newcommand\ours{\textsc{AudioRubrics}\xspace}
\title{Reinforcement Learning with Evolving \\ Rubrics as Rewards for Audio Reasoning}

\author{%
Fangxu Yu\textsuperscript{1}\thanks{Work partially done during the internship at Microsoft Research.},\;\;
Tao Feng\textsuperscript{2},\;\;
Dehai Min\textsuperscript{3},\;\;
Zinan Lin\textsuperscript{4},\;\;
Weijia Xu\textsuperscript{4},\;\; 
\textbf{Michael Xu}\textsuperscript{4},\;\;\\
\textbf{Philip S. Yu}\textsuperscript{\textbf{3}},\;\;
\textbf{Ge Liu}\textsuperscript{\textbf{2}},\;\; 
\textbf{Tianyi Zhou}\textsuperscript{\textbf{5}}
\\
\textsuperscript{1}University of Maryland, College Park,\;\;
\textsuperscript{2}University of Illinois Urbana-Champaign,  \\
\textsuperscript{3}University of Illinois Chicago,\;\;
\textsuperscript{4}Microsoft Research,\;\;
\textsuperscript{5}MBZUAI\\ 
\\
\makebox[\linewidth][c]{%
\huggingface\;\href{https://huggingface.co/collections/umd-zhou-lab/audiorubrics}{Model Weights \& Dataset}\hspace{2em}
\github\;\href{https://github.com/tianyi-lab/AudioRubrics.git}{Code}\hspace{3em}}\\
}

\def\huggingface{\raisebox{-1.5pt}{\includegraphics[height=1.05em]{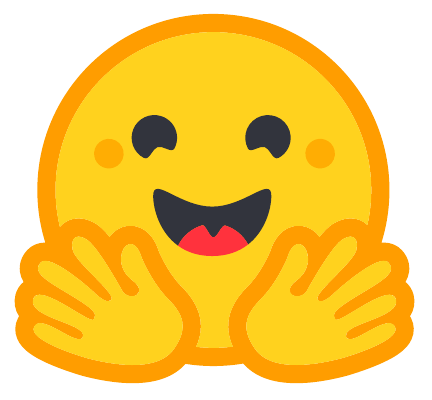}}}
\def\github{\raisebox{-1.5pt}{\includegraphics[height=1.0em]{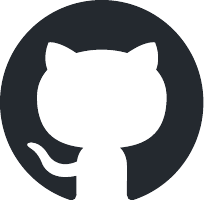}}}

\iclrfinalcopy 
\begin{document}

\maketitle

\begin{abstract}
Audio reasoning is essential for machine understanding of the acoustic world. Reinforcement learning with verifiable rewards can elicit such reasoning, yet existing reward designs are complementary in their limitations: outcome-based rewards supervise only the final answer and let the model reach it without attending to the audio, whereas process-based rewards score the reasoning itself but rely on coarse, hand-crafted, and fixed criteria that neither adapt to each question nor stay grounded in the acoustic evidence. Moreover, questions differ in what they demand, with some hinging on perception and others on multi-step reasoning, and any static criterion weakens as the policy improves. Supervising the reasoning process with fine-grained, audio-grounded, and adaptive rewards is therefore crucial, yet challenging since such rewards are impractical to design by hand for every sample. To this end, we introduce \ours{}, a reinforcement learning framework that supervises audio reasoning with self-evolving, audio-grounded rubric rewards. \ours{} synthesizes per-sample rubrics from the raw waveform and, conditioned on the model's own rollouts, regenerates and reweights criteria per group, supplying a continuous learning signal that keeps targeting the current policy's weaknesses as static criteria saturate. Comprehensive evaluations across three audio reasoning benchmarks reveal that \ours{} substantially outperforms a wide range of open-source and training-based baselines. Furthermore, our analysis shows that the gains scale with the capability of the rubric generator and judge, and \ours{} converges to a stable reasoning length that avoids both degenerate collapse and unbounded growth. The improvement in audio perception further demonstrates the effectiveness of anchoring supervision in the acoustic evidence. Our project page is available at \href{https://audiorubrics.github.io}{https://audiorubrics.github.io}.

\end{abstract}
\section{Introduction}
Audio understanding and reasoning is the ability to perceive acoustic signals and infer meaning from them, spanning the spoken word, environmental sound, and music~\citep{ma2025audiocot, gong2024listen}. As a fundamental facet of human intelligence, this ability lets us extract rich information from what we hear---identifying who is speaking and how, recognizing events from sound alone, and grasping the structure and affect of music---and act on it in the world. Similarly, by equipping machines with the capacity to not merely transcribe audio but reason over its acoustic content, we move beyond shallow pattern recognition toward genuine auditory comprehension, thereby enabling systems that support voice interaction~\citep{chen2025slam} and a broad range of real-world applications from healthcare~\citep{shah2026towards} to multimedia analysis~\citep{polyak2024movie}.

Recently, reinforcement learning with verifiable rewards (RLVR) has emerged as a powerful tool for eliciting reasoning in large audio-language models (LALMs). Outcome-based methods~\citep{li2025reinforcement, rouditchenko2025omni} reward only the final answer. While simple and verifiable, they leave the reasoning trace unsupervised, allowing the model to arrive at the correct option without a genuinely sound reasoning process for understanding the audio. In contrast, process-based methods~\citep{zhifei2025audio, wu2026audio, fan2025incentivizing} score the reasoning itself, providing a denser learning signal, but they rely on coarse and fixed criteria that assess only the surface quality of the reasoning trace. Critically, such criteria do not verify whether the reasoning is \emph{grounded in the audio}, resulting in a response can be rewarded for fluent or well-structured text that is not actually supported by the acoustic evidence.
Beyond this, applying a single fixed rubric to all questions is fundamentally mismatched to the heterogeneity of audio reasoning, since different questions demand different evaluation criteria. Some can be answered from accurate perception alone; for these, soliciting additional reasoning incurs unnecessary computational overhead and increases the risk of hallucination. Others genuinely require multi-step inference, and for these the rubric should place greater weight on the logical soundness of the reasoning chain. A uniform standard rewards both cases identically and therefore fails to match the evaluation to what each question actually requires.
Furthermore, a static rubric becomes progressively less informative as the policy improves. Once the model reliably satisfies a fixed criterion, that criterion no longer discriminates between stronger and weaker responses and its training signal saturates, leaving little gradient to drive further improvement. The evaluation standard thus stops scaling with the model's growing capability.
These limitations call for an evaluation that is multi-dimensional, question-specific rather than uniform, and capability-adaptive, continually raising the bar as the model improves.
However, as summarized in Table~\ref{tab:comparison}, existing approaches cover at most a subset, leaving audio reasoning without an evaluation signal that is audio-grounded, question-adaptive, and capability-scaling.
\begin{table}[t]
    \centering
    \caption{Comparison of \ours with existing post-training designs for audio reasoning. Unlike prior approaches that supervise only the final answer or rely on a coarse, hand-crafted, fixed reward signal, \ours jointly delivers a fine-grained reward that co-evolves with the policy.}
    \label{tab:comparison}
    \setlength{\tabcolsep}{12pt}
    \resizebox{\textwidth}{!}{%
    \begin{tabular}{lccccc}
        \toprule
        \multirow{2}{*}{\textbf{Method}}
        & \makecell{\textbf{Process}\\\textbf{Supervision}}
        & \makecell{\textbf{Fine-grained}\\\textbf{Criteria}}
        & \makecell{\textbf{Audio-}\\\textbf{Grounded}}
        & \makecell{\textbf{Automatic}\\\textbf{Generation}}
        & \makecell{\textbf{Evolve with}\\\textbf{Policy}} \\
        \midrule
        R1-AQA~\citep{li2025reinforcement}
        & \textcolor{red}{\textbf{\ding{55}}} & \textcolor{red}{\textbf{\ding{55}}} & \textcolor{red}{\textbf{\ding{55}}} & \textcolor{red}{\textbf{\ding{55}}} & \textcolor{red}{\textbf{\ding{55}}} \\
        Omni-R1 \citep{rouditchenko2025omni}
        & \textcolor{red}{\textbf{\ding{55}}} & \textcolor{red}{\textbf{\ding{55}}} & \textcolor{red}{\textbf{\ding{55}}} & \textcolor{red}{\textbf{\ding{55}}} & \textcolor{red}{\textbf{\ding{55}}} \\
        Audio-Reasoner \citep{zhifei2025audio}
        & \textcolor{forestgreen}{\textbf{\ding{51}}} & \textcolor{red}{\textbf{\ding{55}}} & \textcolor{forestgreen}{\textbf{\ding{51}}} & \textcolor{red}{\textbf{\ding{55}}} & \textcolor{red}{\textbf{\ding{55}}} \\
        Audio-Thinker \citep{wu2026audio}
        & \textcolor{forestgreen}{\textbf{\ding{51}}} & \textcolor{red}{\textbf{\ding{55}}} & \textcolor{red}{\textbf{\ding{55}}} & \textcolor{forestgreen}{\textbf{\ding{51}}} & \textcolor{red}{\textbf{\ding{55}}} \\
        CESAR \citep{fan2025incentivizing}
        & \textcolor{forestgreen}{\textbf{\ding{51}}} & \textcolor{forestgreen}{\textbf{\ding{51}}} & \textcolor{red}{\textbf{\ding{55}}} & \textcolor{red}{\textbf{\ding{55}}} & \textcolor{red}{\textbf{\ding{55}}} \\
        \midrule
        \rowcolor{cyan!10} \ours
        & \textcolor{forestgreen}{\textbf{\ding{51}}} & \textcolor{forestgreen}{\textbf{\ding{51}}} & \textcolor{forestgreen}{\textbf{\ding{51}}} & \textcolor{forestgreen}{\textbf{\ding{51}}} & \textcolor{forestgreen}{\textbf{\ding{51}}} \\
        \bottomrule
    \end{tabular}}
    \vspace{-20pt}
\end{table}

To address these limitations, we introduce \ours, a reinforcement learning framework that supervises audio reasoning with evolving rubric-based rewards. \ours first constructs a set of initial rubrics for each question in the training data to capture the basic evaluation criteria, generating them directly from the raw waveform so that every criterion is grounded in the acoustic evidence actually present in the clip. Combined with the verifiable RLVR reward, these rubric-based rewards provide dense process-level supervision. Beyond merely checking the final answer, they assess whether the reasoning trace is logically sound and faithfully grounded in the audio, supplying a learning signal on the reasoning process itself.
During RL training, however, the policy steadily improves and satisfies more and more of the initial rubrics. The criteria then become non-discriminative, and the training signal saturates, leaving little gradient to drive further improvement. To resolve this, we condition the rubric generator on the group of rollouts sampled by the policy and compare them against one another. We first prune any rubric whose verdicts are identical across all rollouts, since a criterion that every response passes or fails cannot contribute advantage. By then contrasting where stronger and weaker rollouts diverge, we elicit new criteria that build on the existing ones to promote more advanced reasoning. The rubric set continually retires saturated criteria and adopts harder ones, so the model receives a sustained learning signal whose standard co-evolves with its growing capability.
However, the rubric-based reward incentivizes the model to satisfy more criteria by producing ever longer reasoning chains, which can be exploited to hack the reward through redundant, circular, or hallucinated reasoning. To mitigate this, we incorporate an overthinking penalty that linearly penalizes reasoning length, keeping the trace informative but concise.

We evaluate \ours on three widely used audio reasoning benchmarks: MMAU~\citep{sakshi2024mmau}, MMAR~\citep{ma2025mmar}, and MMSU~\citep{wang2025mmsu}. Experimental results show that \ours consistently outperforms a wide range of open-source models and RL baselines. Further analysis reveals that the rubrics grow progressively more advanced as the policy improves over training, confirming that the evaluation standard co-evolves with the model, and ablation studies verify the contribution of each component.

\begin{figure*}[t]   
\centering
\includegraphics[width=0.94\textwidth]{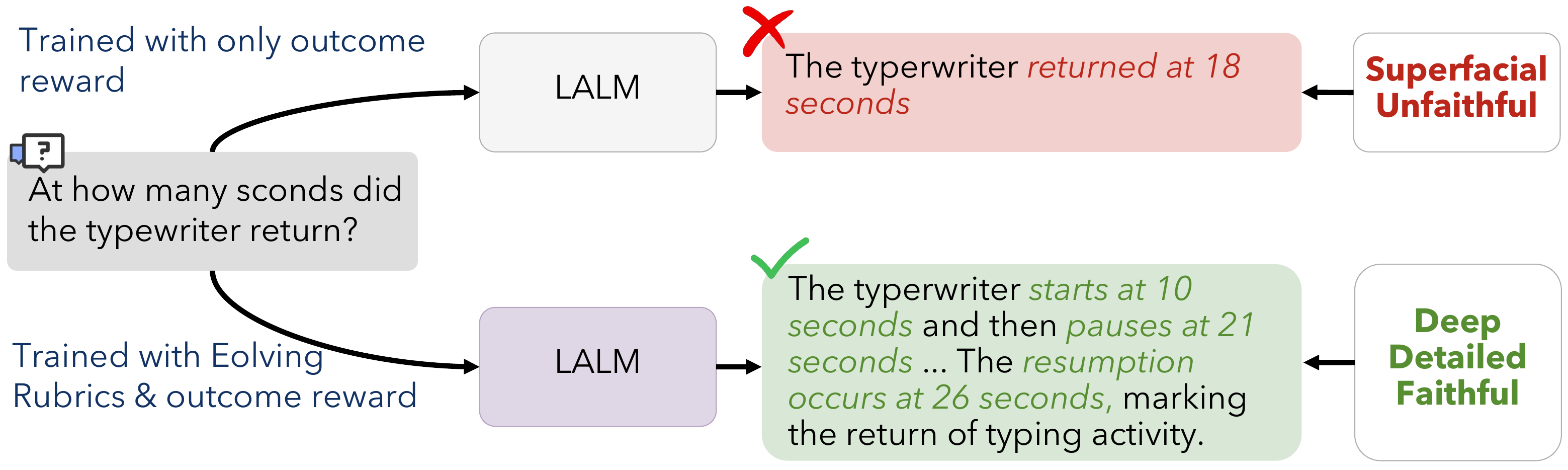}
\caption{Motivation of the proposed \ours. The goal is to progressively enhance the reasoning ability of LALMs by rewarding not only the correct outcome but also the reasoning process with evolving rubrics, which continually distill new criteria from the model's own rollouts to push the reasoning from superficial guesses toward deep, faithful, and evidence-grounded explanations.}
\label{fig:illustration}
\end{figure*}
\section{Preliminary}
\subsection{Group Relative Policy Optimization (GRPO)}
Given a multimodal input ($Q, A$) consisting of an audio input $A$ and a textual query $Q$, the policy model samples $G$ candidate responses $o=\{o_1, \dots,o_G\}$, and a reward function $r(\cdot)$ assigns a reward score to each response, yielding $\{r(o_1), \dots, r(o_G)\}$.
GRPO encourages the LLM to generate responses that maximize a weighted sum reward $R(o)$, defined by:
\begin{equation}
\label{eq:ro}
R(o) = \sum_{i=1}^G \frac{\pi_\theta(o_i)}{\pi_{\theta_{\text{old}}}(o_i)} \cdot \underbrace{\frac{r(o_i) - \text{mean}(\{r(o_j)\}_{j=1}^G)}{\text{std}(\{r(o_j)\}_{j=1}^G)}}_{\text{Advantage } A_i}
\end{equation}

where $\pi_\theta(o)$ denotes the probability of LLM generating the response $o$, and $\pi_{\theta_{\mathrm{old}}}$ represents the policy parameters from a recently optimized state. The latter term is the advantage $A_i$ of the $i$-th candidate.
To ensure training stability and avoid large deviations from the original language model behavior, the final training objective incorporates a KL-divergence regularization term~\citep{guo2025deepseek}, penalizing divergence between $\pi_\theta$ and $\pi_\mathrm{ref}$:
\begin{equation}
\label{eq:grpo}
    \max_{\pi_\theta} \mathbb{E}_{o\sim \pi_{\theta_{\mathrm{old}}}(\cdot \mid Q, A)} [
        R(o) - 
        \beta_{\mathrm{KL}}\, \mathrm{D}_\mathrm{KL}(\pi_\theta \| \pi_\mathrm{ref})
    ]
\end{equation}

where $\beta_{\mathrm{KL}}$ is a scaling coefficient.
We omit the clipping operation for simplicity.

\subsection{Rubric as Rewards}
Rubrics define explicit evaluation criteria for assessing the quality of model responses~\citep{viswanathan2026checklists, gunjal2025rubrics}. We consider sample-wise rubrics, in which the evaluation criteria are specified on a per-example basis in natural language:
given a query $x$ with associated rubrics $\mathcal{R}_x = \{(r_{x,k}, w_{x,k})\}_{k=1}^{K}$, where $r_{x,k}$ denotes a rubric item and $w_{x,k}$ its weight satisfying $\sum_{k=1}^{K} w_{x,k} = 1$, we evaluate a final response $y$ using the rubric-based score
\begin{equation}
S(x, y) = \sum_{k=1}^{K} w_{x,k}\, \textsc{Judge}(x, r_{x,k}, o).
\end{equation}
Each rubric is evaluated by a judge LM that is conditioned on the query $x$ and outputs $\{0, 1\}$ based on whether the response $o$ satisfies $r_{x,k}$.
During training, we optimize the expected rubric score over the training questions using RL.
Using rubrics as rewards offers several advantages: their concrete, well-defined items reduce susceptibility to judge model bias and promote objective evaluation, yielding consistent and comparable scores across different LLM-as-a-judge runs.

\section{\ours: RL with Evolving Rubrics as Rewards}
\label{sec:reward}

\begin{figure*}[t]
\centering
\includegraphics[width=1.0\textwidth]{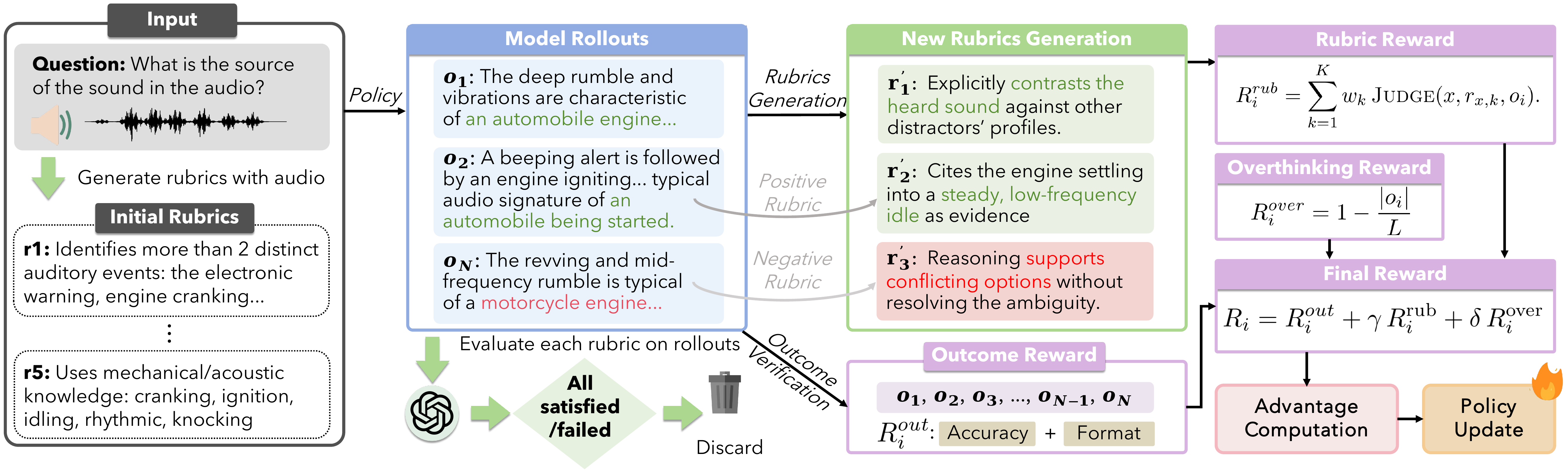}
\caption{Overview of \ours. A rubric generator initializes audio-grounded rubrics for each question. At each step, the policy LALM samples rollouts, and the generator elicits refined rubrics from the rollouts and judges every rubric on them; non-discriminative rubrics are discarded. The resulting rubric score is combined with the outcome reward for GRPO optimization.
}
\label{fig:main_arch}
\end{figure*}

Although RLVR has proven effective in enhancing logical reasoning, it supervises only the final outcome and leaves the reasoning process unchecked. A correct answer certifies neither that the underlying reasoning is grounded in the audio, nor that the reasoning itself is informative (Figure~\ref{fig:illustration}).
Moreover, such a fixed, outcome-level criterion gradually saturates as training progresses, and the learning signal weakens accordingly.
\ours{} addresses both issues by augmenting the outcome reward with an evolving, audio-grounded rubric reward that directly scores the reasoning process.
This design not only encourages high-quality reasoning, but also---because the rubrics evolve alongside the policy---keeps the reward signal informative throughout training.
As shown in Figure~\ref{fig:main_arch}, we first initialize audio-grounded rubrics for each question to assess basic response quality.
At each RLVR step, a judge model evaluates the rollouts against the current rubric set, elicits refined rubrics from the rollouts themselves, prunes non-discriminative ones, and re-weights the survivors; the resulting rubric score is then combined with the outcome reward for GRPO optimization.
\subsection{Task Formulation}
In this work, we investigate the audio reasoning task in the context of large audio language models (LALMs).
Let $\mathcal{D} = (x_1, x_2, \ldots, x_N)$ be an audio reasoning dataset, where each data sample $x_i = (A, Q, y^\star)$ comprises an audio input $A$ (e.g., speech, sound, or music), a textual query $Q$, and the corresponding ground-truth answer $y^\star$.
The audio reasoning task is defined as follows: given a data sample $x_i \in \mathcal{D}$ as input, the LALM is required to reason over the audio and predict an answer $y$ that matches the ground-truth answer $y^\star$.

\subsection{Reward Design}

\noindent\textbf{Outcome Rewards.}
Let $y_i$ be the answer extracted from the \texttt{<answer>} span of rollout $o_i$.
The accuracy reward is $R^{\text{acc}}_i=\mathbb{I}[\,y_i=y^\star\,]$, and the format reward
$R^{\text{fmt}}_i=\mathbb{I}[\,o_i\text{ contains the required }\texttt{<think>}\dots\texttt{</think><answer>}\dots\texttt{</answer>}\text{ structure}\,]$
enforces the reasoning-then-answer template. Both terms are deterministic, and we denote their combination by
\begin{equation}
R^{\text{out}}_i=\alpha\,R^{\text{acc}}_i+\beta\,R^{\text{fmt}}_i,
\label{eq:outcome}
\end{equation}
where $R^{\text{fmt}}_i$ enforces structured output while $R^{\text{acc}}_i$ drives the reasoning capability. 
We refer to RLVR trained with $R_i=R^{\text{out}}_i$ as accuracy-only RLVR.

\noindent\textbf{Audio-Grounded Rubric Initialization.}
Given an audio clip $A$, a question $Q$, and its ground-truth answer $y^\star$, we first synthesize an initial set of weighted rubrics that will act as a process-level reward.
Instead of passing a textual transcript, we feed the raw waveform of $A$ to a single audio-capable model $\Phi$, which serves as both the rubric generator and the judge throughout \ours{}, as native audio input; this keeps every generated criterion anchored to acoustic evidence actually present in the clip.
To spread the rubrics across complementary, non-redundant aspects of response quality, we fix a taxonomy of $K$ evaluation dimensions $\mathcal{F}=\{f_1,\dots,f_K\}$ in advance (e.g., auditory-evidence grounding. See full dimensions in Figure~\ref{fig:prompt-static-rubric}) and request exactly one criterion per dimension,
\begin{equation}
\mathcal{R}_0=\{(r_k,w_k)\}_{k=1}^{K}=\Phi(A,Q,y^\star;\mathcal{F}),
\label{eq:init}
\end{equation}
where each $r_k$ is a binary, positively phrased merit statement and the importance weights satisfy $w_k\in(0,1)$ with $\sum_{k=1}^{K}w_k=1$. The weights are assigned by $\Phi$ in the same call that generates the rubrics.
$\mathcal{R}_0$ is generated once per question before training.

\noindent\textbf{Evolving Rubrics as Rewards.} Though the initial rubrics provide fine-grained criteria, the policy gradually satisfies them as training progresses. Therefore, to obtain a continuous reward signal that targets the weaknesses of the current policy, we design evolving rubrics as rewards. Specifically, at each iteration, conditioned on $A$, $Q$, the current rollouts $\{o_i\}_{i=1}^{G}$, and the rubric set $\mathcal{R}_{\text{prev}}$ retained from the previous iteration in which this question was sampled, $\Phi$ performs three operations. For the first iteration of a question, $\mathcal{R}_{\text{prev}}$ is initialized to the set $\mathcal{R}_0$. Carrying over the survivors from the previous iteration lets the rubric set accumulate the harder criteria distilled in earlier iterations, so the evaluation standard ratchets upward as the policy improves.

\emph{(i) Elicitation and judging.}
$\Phi$ proposes up to $N_{\text{new}}$ new rubrics $\mathcal{R}_{\text{new}}$ that are non-redundant with $\mathcal{R}_{\text{prev}}$ and target reasoning quality and audio grounding.
Because these rubrics are induced from observed rollouts, they also carry negative criteria (satisfied is a flaw), capturing recurring failure modes that positively phrased initial rubrics cannot express.
$\Phi$ then returns binary judgments $j_{k,i}\in\{0,1\}$ of whether rollout $o_i$ satisfies rubric $r_k$ for every $r_k\in\mathcal{R}=\mathcal{R}_{\text{prev}}\cup\mathcal{R}_{\text{new}}$ and every rollout.
Polarity normalization flips negative-polarity judgments so that $b_{k,i}=1$ always denotes a good outcome:
\begin{equation}
b_{k,i}=
\begin{cases}
j_{k,i}, & r_k\ \text{is positive rubric},\\[2pt]
1-j_{k,i}, & r_k\ \text{is negative rubric},
\end{cases}
\label{eq:polarity}
\end{equation}
yielding comparable binary vectors $\{b_{k,i}\}_{i=1}^{G}$ across rubrics.

\emph{(ii) Variance filtering.}
For each rubric, we compute the within-group standard deviation $s_k=\operatorname{std}_i\{b_{k,i}\}_{i=1}^{G}$ and prune every rubric with $s_k=0$.
Such a rubric is either satisfied by all rollouts, indicating that the policy has already mastered this criterion, or by none, indicating that it lies beyond the policy's current capability. In both cases, it assigns every rollout the same verdict, and thus contributes nothing to the group-relative advantage. Among the survivors, we rank rubrics by $s_k$ and retain the top $M$ most discriminative ones or all, if fewer than $M$ survive. We denote the kept set of rubrics by $\mathcal{K}$.

\emph{(iii) Weighting and scoring.}
Conditioned on $A$, $Q$, and $\mathcal{K}$, $\Phi$ re-assigns positive weights $\{w_k\}_{k\in\mathcal{K}}$ with $\sum_{k\in\mathcal{K}}w_k=1$. The weights are re-assigned by $\Phi$ for this question, and the final rubric reward is:
\begin{equation}
R^{\text{rub}}_i=\sum_{k\in\mathcal{K}}w_k\,b_{k,i}\ \in[0,1].
\label{eq:rubric-reward}
\end{equation}
This elicitation, filtering, and re-weighting is performed at every training step, independently for each prompt group. Because rubrics are thus updated per group conditioned on the actual rollouts, $R^{\text{rub}}_i$ stays informative and provides a more advanced reward signal for the current policy.

\noindent\textbf{Overthinking Penalty.} Although the rubric reward improves reasoning quality, satisfying more criteria favors longer reasoning, which risks degenerating into redundant or circular traces that accumulate errors and hallucinations~\citep{mahmoud2026reward}.
To alleviate this, we regularize reasoning length with a penalty:
\begin{equation}
R^{\text{over}}_i=1-\frac{|o_i|}{L},
\label{eq:over}
\end{equation}
where $|o_i|$ is the token length of the reasoning trace $o_i$ and $L$ is a reference length that normalizes the penalty and sets its scale relative to the other reward terms.
Taken in isolation this term favors shorter traces, but in combination it counterbalances the rubric reward, which pushes toward longer traces that satisfy more criteria. The rubric term thus improves reasoning quality with richer details, while the length term bounds its quantity, and their equilibrium keeps the trace informative but concise.

\noindent\textbf{Overall Reward and Optimization.}
The three reward terms above play complementary roles, in which the outcome reward anchors final-answer correctness, the rubric reward densely supervises the reasoning process, and the overthinking penalty bounds its length. The per-rollout reward is their weighted combination:
\begin{equation}
R_i=R^{\text{out}}_i+\gamma\,R^{\text{rub}}_i+\delta\,R^{\text{over}}_i,
\label{eq:reward}
\end{equation}
where $\gamma$ controls the impact of the rubric reward and $\delta$ weights the overthinking penalty. Accuracy-only RLVR is recovered at $\gamma=\delta=0$.
During training, we replace the reward function $r(\cdot)$ in Eq.~\ref{eq:ro} with our evolving rubric reward $R_i$ and train the LALM to maximize the GRPO objective in Eq.~\ref{eq:grpo}. Algorithm~\ref{alg:rubric} in the appendix summarizes the per-group computation of the evolving rubric reward.

\section{Experiments}
\subsection{Experimental Setups}
\noindent\textbf{Benchmarks.}
We evaluate on three audio reasoning benchmarks. MMAU Test-mini~\citep{sakshi2024mmau} has 1000 multiple-choice questions over 27 tasks spanning speech, sound, and music. MMAR~\citep{ma2025mmar} provides 1000 QA pairs from real-world videos that mix the three modalities, organized into four reasoning layers for deep reasoning. MMSU~\citep{wang2025mmsu} adds 5000 triplets over 47 spoken-language tasks probing fine-grained paralinguistic and phonological cues such as prosody, emotion, and speaker traits. Together they cover sound, music, and speech and their mixtures, from perception to multi-step reasoning.

\noindent\textbf{Baselines and Evaluation Metrics.} We compare \ours against a comprehensive suite of baselines spanning
proprietary models, open-source LALMs,
and recent training-based audio reasoning methods, including 3
proprietary models and 13 open-source models. For proprietary models,
we evaluate GPT-4o-Audio~\citep{openai2024gpt4oaudio},
GPT-audio-1.5~\citep{openai2026gptaudio15}, and
Gemini-3-Flash, Gemini-3.1-Pro~\citep{deepmind2025gemini31pro}.
Open-source LALMs include
Qwen2-Audio~(7B)~\citep{chu2024qwen2},
Phi-4-Multimodal~(5.6B)~\citep{abdin2024phi},
Kimi-Audio~(7B)~\citep{ding2025kimi},
Step-Audio-2-mini~(7B)~\citep{wu2025step},
Qwen2.5-Omni~(7B)~\citep{xu2025qwen25omni},
MiMo-Audio~(7B)~\citep{zhang2025mimo}, and
Audio-Flamingo~3~(7B)~\citep{ghosh2026audio}. For
training-based reasoning methods, we further compare with
Audio-Reasoner~(7B)~\citep{zhifei2025audio},
R1-AQA~(7B)~\citep{li2025reinforcement},
Omni-R1~(7B)~\citep{rouditchenko2025omni},
Ke-Omni-R~(7B)~\citep{ke_omni_r},
Audio-Thinker~(7B)~\citep{wu2026audio}, and
CESAR~(7B)~\citep{fan2025incentivizing}, all of which apply RL or SFT-based post-training to elicit audio reasoning. For evaluation, we set temperature $\tau=0$ to perform greedy decoding for all models and use accuracy as the main evaluation metric.

\noindent\textbf{Training details.} The training data is drawn from the AVQA dataset~\citep{yang2022avqa}, which is also used by most training-based baselines
(R1-AQA, Omni-R1, Ke-Omni-R, Audio-Thinker, and CESAR). Following R1-AQA~\citep{li2025reinforcement}, we extract audio from videos and construct audio-text pairs by replacing “video” with “audio” in the questions, resulting in 40,176 training samples. All training runs use 4 H100 GPUs.  We use Gemini-3.1-Pro as the rubric judge and generator. Our experiments employ Qwen2.5-Omni-7B, sampling $G=8$ responses per training example. See more implementation details in Appendix~\ref{sec: implementation}.

\begin{table*}[t]
\centering
\caption{Results on the MMSU and MMAU Test-mini benchmarks. Top two results are highlighted in \textbf{bold} and \underline{underline}, respectively. Audio-Thinker is not evaluated on MMSU as its checkpoint is not publicly released.}
\label{tab:mmsu_mmau}
\resizebox{\textwidth}{!}{%
\begin{tabular}{l|ccccccccc|cccc}
\toprule
\multirow{3}{*}{\textbf{Models}}
 & \multicolumn{9}{c}{\cellcolor{gMMSU}\textbf{MMSU (\%$\uparrow$)}}
 & \multicolumn{4}{c}{\cellcolor{gMMAU}\textbf{MMAU Test-mini (\%$\uparrow$)}} \\
\cmidrule(lr){2-10} \cmidrule(lr){11-14}
 & \multicolumn{4}{c}{\textbf{Perception}} & \multicolumn{4}{c}{\textbf{Reasoning}}
 & \multirow{2}{*}{\textbf{Avg}}
 & \multirow{2}{*}{\textbf{Sound}} & \multirow{2}{*}{\textbf{Music}}
 & \multirow{2}{*}{\textbf{Speech}} & \multirow{2}{*}{\textbf{Avg}} \\
\cmidrule(lr){2-5} \cmidrule(lr){6-9}
 & Seman. & Phono. & Para. & Avg & Seman. & Phono. & Para. & Avg & All & & & & \\
\midrule
\rowcolor{barGray}\multicolumn{14}{c}{\textit{Proprietary Models}} \\\midrule
GPT-4o-Audio     & 59.70 & 41.56 & 21.44 & 39.67 & 80.83 & 78.74 & 26.25 & 71.96 & 56.38 & 64.56 & 56.29 & 66.67 & 62.50 \\
GPT-audio-1.5    & 82.20 & 64.06 & 29.11 & 54.84 & 90.61 & 65.71 & 32.54 & 72.52 & 63.40 & 78.08 & 63.17 & 83.48 & 74.90 \\
Gemini-3-Flash  & 80.94 & 68.98 & 65.45 & 70.54 & 92.06 & 90.38 & 57.01 & 86.53 & 78.28 & 74.47 & 76.65 & 81.38 & 77.50 \\
Gemini-3.1-Pro & 90.08 & 76.79 & 79.41 & 81.09 & 88.36 & 87.41 & 61.19 & 84.21 & 82.60 & 75.98 & 75.45 & 81.38 & 77.60 \\
\midrule
\rowcolor{barGray}\multicolumn{14}{c}{\textit{Open-source Models}} \\\midrule
Qwen2-Audio      & 52.14 & 32.87 & 35.56 & 39.02 & 77.62 & 64.81 & 46.67 & 68.90 & 53.27 & 67.27 & 56.29 & 55.26 & 59.60 \\
Phi-4-Multimodal & 38.72 & 34.86 & 29.56 & 33.41 & 57.81 & 65.94 & 42.09 & 57.59 & 44.96 & 65.47 & 64.37 & 67.27 & 65.70 \\
Kimi-Audio       & 57.64 & 42.30 & 35.74 & 43.52 & 81.77 & 76.65 & \underline{55.22} & 76.03 & 59.28 & 75.68 & 66.77 & 62.16 & 68.20 \\
Step-Audio-2-mini& 38.90 & 31.55 & 41.19 & 37.13 & 70.04 & 71.44 & 44.18 & 67.02 & 51.60 & 79.30 & 68.44 & 68.16 & 72.73 \\
Qwen2.5-Omni-7B  & 55.12 & 37.33 & 39.35 & 42.50 & 88.00 & \textbf{81.37} & 48.36 & \underline{79.83} & 60.57 & 69.07 & 59.58 & 66.97 & 65.20 \\
MiMo-Audio-7B    & 57.80 & 43.74 & 38.22 & 45.04 & 73.74 & 66.73 & 38.21 & 65.99 & 55.18 & 68.47 & \textbf{82.58} & 73.65 & 74.90 \\
Audio-Flamingo 3 & -- & -- & -- & -- & -- & -- & -- & -- & 62.30 & 79.88 & \underline{76.55} & 66.37 & 74.26 \\
\midrule
\rowcolor{barGray}\multicolumn{14}{c}{\textit{Training-based models}} \\\midrule
Audio-Reasoner & 46.77 & 35.19 & 34.16 & 37.64 & 75.00 & 64.38 & 37.01 & 65.45 & 51.10 & 67.87 & 69.16 & 66.07 & 67.70 \\
R1-AQA         & 55.12 & 35.19 & 37.72 & 41.09 & 79.15 & 66.22 & 51.04 & 70.04 & 55.10 & 68.77 & 64.37 & 63.66 & 65.60 \\
Omni-R1        & 58.43 & 44.49 & \underline{42.48} & 47.13 & 88.18 & 77.79 & 46.87 & 78.26 & 62.20 & 81.38 & 68.86 & 73.57 & 74.60 \\
Ke-Omni-R      & 58.74 & 46.31 & 40.50 & 47.09 & 86.82 & 74.31 & \textbf{60.00} & 78.06 & 62.08 & 79.28 & 70.06 & 74.47 & 74.60 \\
Audio-Thinker  & -- & -- & -- & -- & -- & -- & -- & -- & -- & 77.48 & 70.36 & 73.37 & 73.70 \\
CESAR          & 60.16 & \underline{50.16} & 39.50 & \underline{48.45} & \underline{88.72} & 80.66 & 57.01 & \textbf{81.07} & \underline{64.24} & \underline{83.48} & 73.05 & 74.77 & \underline{77.10} \\\midrule
\ours          & \textbf{66.14} & \textbf{51.34} & \textbf{44.16} & \textbf{52.75} & \textbf{89.17} & \underline{81.26} & 49.25 & 80.45 & \textbf{65.86} & \textbf{85.89} & 72.16 & \textbf{75.98} & \textbf{78.00} \\
\bottomrule
\end{tabular}%
}
\vspace{-5pt}
\end{table*}

\subsection{Main Results}
We present the performance comparison between \ours and existing powerful models and methods across three widely-used audio understanding and reasoning benchmarks in Tables~\ref{tab:mmsu_mmau} and~\ref{tab:mmar}. Based on the results, we have the following key observations:

\textbf{(i) \ours{} achieves superior performance on all benchmarks among all models and methods of similar size.}
\ours{} attains the best overall accuracy on all three benchmarks.
The advantage is also evident at the dimension level. \ours{} ranks first on the Sound category of both MMAU and MMAR across all evaluated models, and leads all similar-size models on the Speech category of both benchmarks. \ours also leads or matches the best similar-size baseline on three of the four mixed-modality splits of MMAR. These consistent gains at both the overall and the per-dimension level demonstrate the effectiveness of our approach, which provides a richer learning signal for training.

\textbf{(ii) The audio perception capability of \ours{} shows tangible improvements.} Specifically, on the perception split of MMSU, \ours{} surpasses the best-performing baseline of comparable size by 4.3 points, an 8.9\% relative improvement, with consistent gains across all three perception dimensions. This indicates that our approach teaches the model to anchor its reasoning in the acoustic evidence it actually hears, thereby translating process-level supervision into a genuine gain in perception and ensuring that correct answers rest on faithfully perceived evidence.
\begin{figure*}[t]   
\centering
\includegraphics[width=1.0\textwidth]{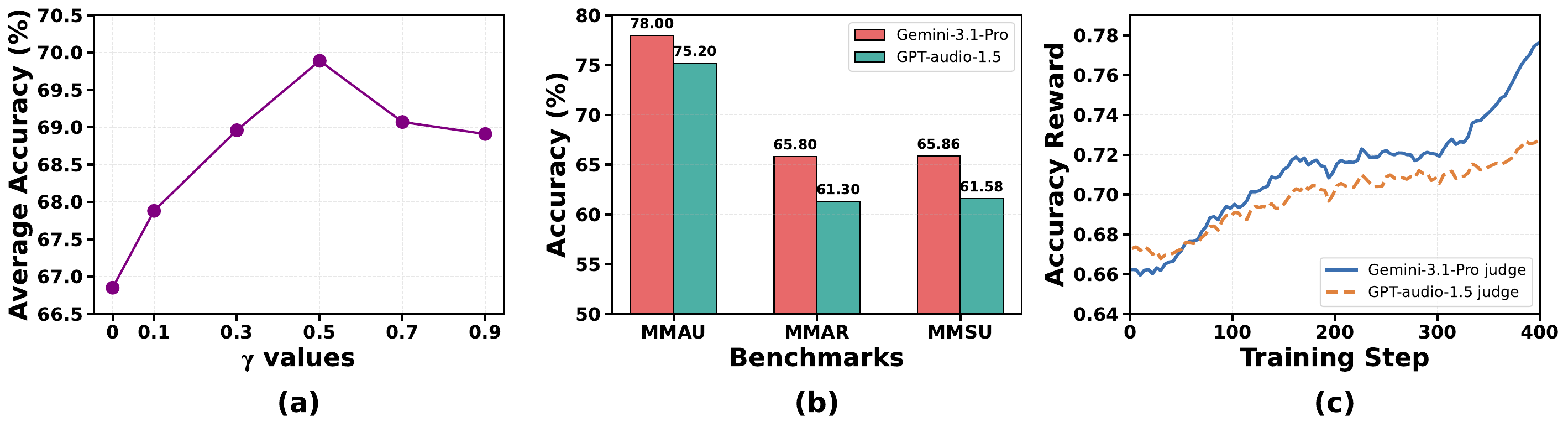}
\caption{(a) Average performance across all benchmarks with varying rubric reward weight $\gamma$. (b) Comparison of accuracy across benchmarks when using different judging and reward generators. (c) Dynamics of accuracy reward during training when using different judging and reward generators.
}
\label{fig:rubric_analysis}
\vspace{-13pt}
\end{figure*}

\begin{table*}[t]
\centering
\caption{Results on the MMAR benchmark. Top two results are highlighted in \textbf{bold} and \underline{underline}.}
\label{tab:mmar}
\resizebox{\textwidth}{!}{
\begin{tabular}{l|ccc|cccc|c}
\toprule
\multirow{2}{*}{\textbf{Method}}
 & \multicolumn{3}{c|}{\cellcolor{gMMAU}\textbf{Single Modality}}
 & \multicolumn{4}{c|}{\cellcolor{gMMAR}\textbf{Mixed Modality}}
 & \cellcolor{gMMSU}\textbf{Overall} \\
\cmidrule(lr){2-4}\cmidrule(lr){5-8}
 & \textbf{Sound} & \textbf{Music} & \textbf{Speech}
 & \textbf{Mix-S-M} & \textbf{Mix-S-Sp} & \textbf{Mix-M-Sp} & \textbf{Mix-S-M-Sp}
 & \\
\midrule
\rowcolor{barGray}\multicolumn{9}{c}{\textit{Proprietary Models}} \\
\midrule
GPT-4o Audio & 53.94 & 50.97 & 70.41 & 63.64 & 72.48 & 62.20 & 75.00 & 63.50 \\
GPT-audio-1.5 & 63.64 & 53.40 & 81.29 & 63.64 & 83.49 & 73.17 & 70.83 & 72.00 \\
Gemini-3-Flash & 66.06 & 62.62 & 85.71 & 90.91 & 83.03 & 75.61 & 75.00 & 76.10 \\
Gemini-3.1-Pro & 70.91 & 65.53 & 84.01 & 90.91 & 84.86 & 79.27 & 66.67 & 77.50 \\
\midrule
\rowcolor{barGray}\multicolumn{9}{c}{\textit{Large Audio Language Models (LALMs)}} \\
\midrule
Qwen2-Audio & 33.33 & 24.27 & 32.31 & 9.09 & 31.19 & 30.49 & 25.00 & 30.00 \\
Phi-4-multimodal & 32.73 & 31.07 & 50.68 & 18.18 & 41.74 & 57.32 & 41.67 & 41.70 \\
Kimi-Audio  & -- & -- & -- & -- & -- & -- & --  & 48.00 \\
Step-Audio-2-mini & 51.52 & 32.04 & 49.66 & 45.45 & 51.83 & 54.88 & 62.50 & 47.50 \\
Qwen2.5-Omni-7B & 58.79 & 40.78 & 59.86 & 54.55 & 61.93 & \underline{67.07} & 58.33 & 56.70 \\
MiMo-Audio-7B & 52.73 & 37.38 & 62.59 & 36.36 & 69.27 & 59.76 & 62.50 & 56.70 \\
Audio-Flamingo 3 & -- & -- & -- & -- & -- & -- & --  & 60.10 \\
\midrule
\rowcolor{barGray}\multicolumn{9}{c}{\textit{Training-based models}} \\
\midrule
Audio-Reasoner & 43.64 & 33.50 & 32.99 & 45.45 & 42.66 & 31.71 & 25.00 & 36.80 \\
R1-AQA & 52.73 & 40.78 & 49.32 & 9.09 & 50.92 & 52.44 & 50.00 & 48.30 \\
Omni-R1 & 59.39 & 50.49 & 61.56 & 54.55 & 60.09 & 63.41 & 41.67 & 58.20 \\
Ke-Omni-R & 63.64 & 47.09 & 62.93 & \underline{63.64} & 68.35 & \underline{67.07} & 45.83 & 60.90 \\
Audio-Thinker & \underline{67.27} & \underline{53.88} & \underline{64.29} & \textbf{72.73} & \underline{71.56} & \textbf{73.17} & \underline{66.67} & \underline{65.30} \\
CESAR & 66.06 & \textbf{55.83} & 62.24 & \underline{63.64} & 67.43 & 60.98 & \underline{66.67} & 62.70 \\
\midrule
\ours & \textbf{68.48} & 51.94 & \textbf{68.71} & \textbf{72.73} & \textbf{72.02} & 65.85 & \textbf{70.83} & \textbf{65.80} \\
\bottomrule
\end{tabular}
}
\vspace{-10pt}
\end{table*}
\subsection{Further Analysis on Reward Design and Judgement}

\begin{wraptable}{r}{0.55\textwidth}
\vspace{-12pt}
\centering
\small
\setlength{\tabcolsep}{4pt}
\caption{\textbf{Impact of rubric reward weighting.} Performance on the audio
benchmarks with different rubric weights.
$\Delta^{\%}_{rel}$ is relative to the GRPO baseline.}
\label{tab: rubric_weight}
\begin{tabular}{l|cc|cc|cc}
\toprule
\multirow{2}{*}{\textbf{Rubric}}
& \multicolumn{2}{c|}{\cellcolor{gMMAU}\textbf{MMAU}}
& \multicolumn{2}{c|}{\cellcolor{gMMAR}\textbf{MMAR}}
& \multicolumn{2}{c}{\cellcolor{gMMSU}\textbf{MMSU}} \\
& \textbf{ACC} & $\Delta^{\%}_{rel}$
& \textbf{ACC} & $\Delta^{\%}_{rel}$
& \textbf{ACC} & $\Delta^{\%}_{rel}$ \\
\midrule
GRPO & 75.20 & -- & 62.20 & -- & 63.14 & -- \\
\midrule
$\gamma=0.10$ & 77.10 & \cellcolor{green!18}\up{2.53}
     & 62.80 & \cellcolor{green!7}\up{0.96}
     & 63.74 & \cellcolor{green!7}\up{0.95} \\
$\gamma=0.30$ & 77.30 & \cellcolor{green!20}\up{2.79}
     & 64.90 & \cellcolor{green!30}\up{4.34}
     & 64.68 & \cellcolor{green!18}\up{2.44} \\
$\gamma=0.50$& \textbf{78.00} & \cellcolor{green!26}\up{3.72}
     & \textbf{65.80} & \cellcolor{green!36}\up{5.79}
     & 65.86 & \cellcolor{green!30}\up{4.31} \\
$\gamma=0.70$ & 77.70 & \cellcolor{green!24}\up{3.32}
     & 63.40 & \cellcolor{green!13}\up{1.93}
     & \textbf{66.12} & \cellcolor{green!33}\up{4.72} \\
$\gamma=0.90$ & 77.80 & \cellcolor{green!24}\up{3.46}
     & 64.10 & \cellcolor{green!22}\up{3.05}
     & 64.84 & \cellcolor{green!20}\up{2.69} \\
\bottomrule
\end{tabular}
\vspace{-10pt}
\end{wraptable}

\textbf{Effect of the rubric reward weight.} Table~\ref{tab: rubric_weight} and Figure~\ref{fig:rubric_analysis}\,(a) vary the rubric weight $\gamma$ with the overthinking penalty fixed. Across the entire range, adding the rubric reward consistently improves over GRPO. As $\gamma$ increases, \ours{} receives a stronger process-level reward and its accuracy improves accordingly, peaking at $\gamma=0.5$. Beyond this point, however, performance declines since an overly large rubric weight downweights the accuracy reward, so the policy is increasingly optimized toward satisfying the rubric criteria rather than producing the correct answer, ultimately weakening the very signal that grounds the reasoning in a verifiably correct outcome. 

\begin{wraptable}{r}{0.55\textwidth}
\centering
\small
\vspace{-12pt}
\setlength{\tabcolsep}{4pt}
\caption{\textbf{Impact of overthinking penalty weighting.} Performance on the audio
benchmarks with different overthinking penalty weights. $\Delta^{\%}_{rel}$ is relative to the GRPO baseline.}
\label{tab:ot_weight}
\begin{tabular}{l|cc|cc|cc}
\toprule
\multirow{2}{*}{\textbf{OT}}
& \multicolumn{2}{c|}{\cellcolor{gMMAU}\textbf{MMAU}}
& \multicolumn{2}{c|}{\cellcolor{gMMAR}\textbf{MMAR}}
& \multicolumn{2}{c}{\cellcolor{gMMSU}\textbf{MMSU}} \\
& \textbf{ACC} & $\Delta^{\%}_{rel}$
& \textbf{ACC} & $\Delta^{\%}_{rel}$
& \textbf{ACC} & $\Delta^{\%}_{rel}$ \\
\midrule
GRPO & 75.20 & -- & 62.20 & -- & 63.14 & -- \\
\midrule
$\delta=0.05$ & 77.40 & \cellcolor{green!21}\up{2.93}
     & 63.70 & \cellcolor{green!17}\up{2.41}
     & 64.30 & \cellcolor{green!13}\up{1.84} \\
$\delta=0.10$ & 77.70 & \cellcolor{green!23}\up{3.32}
     & \textbf{66.10} & \cellcolor{green!44}\up{6.27}
     & 65.56 & \cellcolor{green!27}\up{3.83} \\
$\delta=0.15$ & \textbf{78.00} & \cellcolor{green!26}\up{3.72}
     & 65.80 & \cellcolor{green!41}\up{5.79}
     & \textbf{65.86} & \cellcolor{green!30}\up{4.31} \\
$\delta=0.20$ & 77.20 & \cellcolor{green!19}\up{2.66}
     & 64.40 & \cellcolor{green!25}\up{3.54}
     & 64.70 & \cellcolor{green!17}\up{2.47} \\
\bottomrule
\end{tabular}
\vspace{-10pt}
\end{wraptable}

\textbf{Effect of the overthinking penalty.} Table~\ref{tab:ot_weight} studies the length-penalty weight $\delta$. When the weight $\delta$ is small, the policy tends to generate overlong reasoning traces, since it seeks to satisfy more rubrics, which weakens the impact of the accuracy reward. As $\delta$ increases, the performance peaks at $\delta=0.15$, which effectively controls overthinking. However, a larger penalty would excessively compress the reasoning trace, weakening reasoning ability and leading to suboptimal performance. See Appendix~\ref{sec:training_dynamics} for the length dynamics during training.

\textbf{Effect of the Rubric Generator and Judge Model.} To examine how the choice of generator and judge affects \ours{}, we replace Gemini-3.1-Pro with the weaker GPT-audio-1.5 as both. As shown in Figure~\ref{fig:rubric_analysis}\,(b), performance drops with GPT-audio-1.5, even falling below the vanilla GRPO baseline. The training dynamics in Figure~\ref{fig:rubric_analysis}\,(c) corroborate this: after a brief warmup in which the two runs are comparable, the Gemini-3.1-Pro policy pulls ahead and stays superior, whereas the GPT-audio-1.5 variant learns more slowly and plateaus lower. We attribute this gap to the lower-quality rubrics from the weaker model, whose imprecise or poorly grounded criteria yield noisy rewards, indicating that \ours{} hinges on a sufficiently capable generator and judge.

\subsection{Ablation Study}
We conduct ablation studies to analyze the contribution of each component in \ours. As shown in Table~\ref{tab:ablation} Compared to the base model, RL training substantially improves audio reasoning capability, confirming its necessity. Building on RL training, the introduction of rubric rewards yields a further gain, where the initial rubrics provide fine-grained criteria that jointly assess the quality of the reasoning process and the accuracy of audio perception. Allowing the rubrics to evolve, rather than relying solely on their initialized form, brings additional improvement, as the evolving rubrics continually target the current model's reasoning weaknesses and supply progressively more advanced criteria that drive further gains in reasoning ability. The overthinking penalty mitigates the tendency to generate excessively long reasoning traces merely to satisfy the rubrics at the expense of answer accuracy. Overall, the ablation demonstrates that each component contributes a meaningful performance gain while functioning in concert as an integrated whole.

\begin{table*}[t]
\centering
\caption{Ablation results of different components in \ours.}
\vspace{-5pt}
\label{tab:ablation}
\resizebox{\textwidth}{!}{
\begin{tabular}{l|cccc|ccc}
\toprule
\multirow{2}{*}{\textbf{Method}}
 & \multicolumn{4}{c|}{\cellcolor{gMMAU}\textbf{Components}}
 & \multicolumn{3}{c}{\cellcolor{gMMAR}\textbf{Overall Accuracy (\%)}} \\
\cmidrule(lr){2-5} \cmidrule(lr){6-8}
 & \textbf{RL Training} & \textbf{Static Rubrics} & \textbf{Evolving Rubrics} & \textbf{Length Penalty}
 & \textbf{MMAU} & \textbf{MMAR} & \textbf{MMSU} \\
\midrule
\textbf{Full Method (\ours)}            & \cmark & \cmark & \cmark & \cmark & 78.00 & 65.80 & 65.86 \\\midrule
\quad Ablating Length Penalty           & \cmark & \cmark & \cmark & \xmark & 77.20 & 64.60 & 65.22 \\
\quad Ablating Evolving Rubrics         & \cmark & \cmark & \xmark & \xmark & 76.20 & 63.60 & 65.44 \\
\quad Ablating Static Rubrics           & \cmark & \xmark & \xmark & \xmark & 75.20 & 62.20 & 63.14 \\
\quad Ablating RL Training & \xmark & \xmark & \xmark & \xmark & 65.20 & 56.70 & 60.57\\
\bottomrule
\end{tabular}
}
\vspace{-15pt}
\end{table*}

\subsection{Rubrics Evolving Analysis}
\begin{wrapfigure}{r}{0.37\textwidth}
    \vspace{-15pt}
	\centering
	\includegraphics[width=0.9\linewidth]{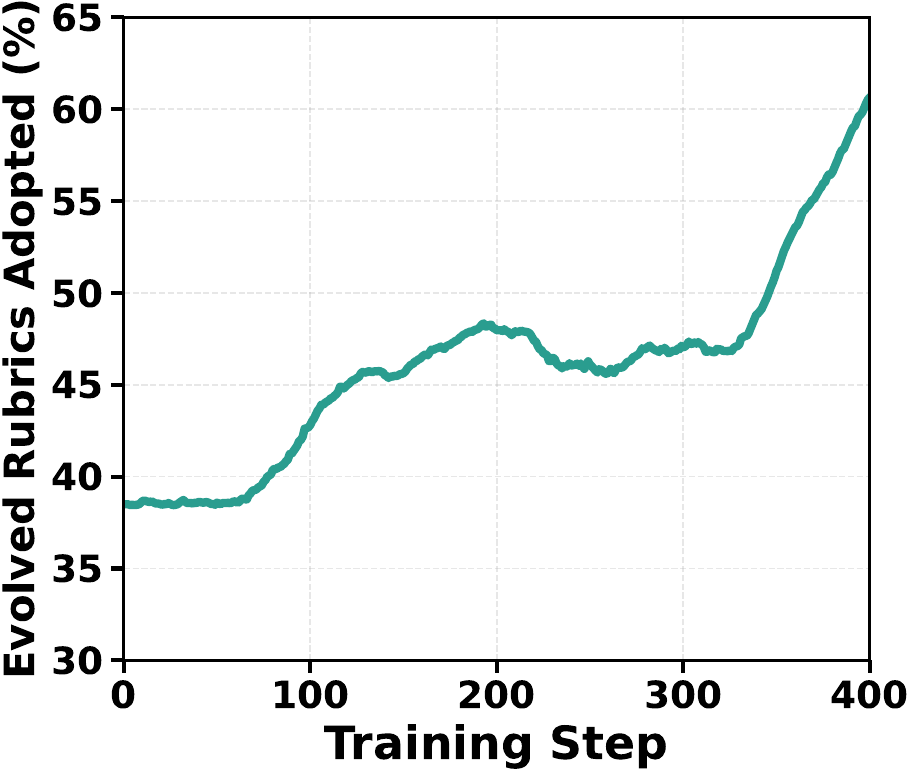}
	\caption{The ratio of newly evolved rubrics adopted during training.}
    \label{fig:evolving_ratio}
    \vspace{-5pt}
	\vspace{-10pt}
\end{wrapfigure}
To understand how the rubric set adapts during training, we track the ratio of newly evolved rubrics that are adopted for the reward at each step. As shown in Figure~\ref{fig:evolving_ratio}, this share grows steadily from roughly 35\% early in training to about 60\% by the end. In the early stage, the policy mainly satisfies the initial rubrics, so few new criteria are needed. As the model improves and satisfies the easier rubrics, the variance filter increasingly prunes them, and a growing fraction of the supervision comes from newly generated rubrics that provide more advanced, fine-grained criteria. These results demonstrate that the rubric set continually upgrades to track the model's current weaknesses, rather than supervising against a frozen rubric list. See Appendix~\ref{sec: case rubrics} for a qualitative analysis.

\section{Related Work}
\textbf{Audio Understanding and Reasoning.} Early work learns audio representations by adapting CLIP to audio~\citep{radford2021learning, wu2022wav2clip}. More recent Large Audio Language Models (LALMs) integrate audio into large language model backbones and, after large-scale pretraining~\citep{fla2, qwen1_audio, xu2025qwen25omni, chu2024qwen2}, sharpen reasoning through supervised fine-tuning on reasoning traces~\citep{zhifei2025audio, li2026audio}, tool use~\citep{chen2026audiorouter, tong2026autagent, lee2025audio}, or reinforcement learning~\citep{fan2025incentivizing, tian2025step, wu2026audio, yu2026weak}. Unlike these methods, which reward only the final answer or score reasoning against fixed, hand-crafted criteria, \ours{} supervises audio reasoning with evolving, audio-grounded rubrics tailored to each question.

\textbf{Rubric as Rewards.}
Reinforcement learning with verifiable rewards excels where correctness is programmatically checkable~\citep{guo2025deepseek, yu2024flow, yu2026arrowgev, yu2026dapo}, but its scalar reward does not extend to open-ended tasks that demand multi-criteria judgment. Rubric-based rewards fill this gap by scoring responses against fine-grained, instance-specific criteria~\citep{bi2025reward, raghavendra2026agentic, mu2024rule, zhou2025breaking, tyagi2026not}, which are elicited from a strong model and synthesized either offline or online from policy rollouts~\citep{kim2024prometheus, gupta2025carmo, liu2025openrubrics, xie2025auto, shen2026rethinking, shao2025dr, sheng2026reinforcing, rezaei2025online, xu2026alternating}, and have recently been extended to vision-language and omni-modal reasoning~\citep{jia2025autorubric, yu2026visual, chen2026rucl, qiu2026rationale, kong2026omni}. Different from these text- and vision-centric efforts, we introduce evolving rubrics to audio reasoning, providing a more fine-grained and adaptive learning signal for training LALMs to reason.
\vspace{-5pt}
\section{Conclusion}
\vspace{-5pt}
We introduce \ours{}, a reinforcement learning framework that supervises audio reasoning with rubric-based rewards. To sustain the learning signal as the policy improves, we further evolve the rubrics from the model's own rollouts and elicit harder ones, fostering an evaluation standard that co-evolves with the model's capability. Extensive experiments demonstrate that \ours{} substantially outperforms a wide range of baselines on audio understanding and reasoning benchmarks.

\bibliography{iclr2026_conference}
\bibliographystyle{iclr2026_conference}
\appendix
\newpage
\section{Implementation Details}
\label{sec: implementation}
We leverage Qwen2.5-Omni-7B~\citep{xu2025qwen25omni} as our base model. During RLVR training, we optimize the policy with GRPO for 400 steps, and set a batch size of 8, learning rate 1e-6 with a constant schedule, number of rollouts $G=8$ with sampling temperature 1.0, and KL coefficient $\beta_{\mathrm{KL}}=0.001$, with a maximum prompt length of 4096 and a maximum response length of 1024 tokens under bf16 precision. For the reward, we set the accuracy and format weights to $\alpha=0.9$ and $\beta=0.1$, the rubric reward weight to $\gamma=0.5$ and the overthinking penalty weight to $\delta=0.15$, following the sensitivity analyses in Section~4.3, and the target reasoning length to $L=256$ tokens. Gemini-3.1-Pro serves as the rubric generator and judge $\Phi$, keeping the top $M=5$ most discriminative rubrics per group and eliciting up to $N_{\text{new}}=3$ new rubrics at each step. The checkpoint from the final step is used for all evaluations. All experiments were conducted on 4 H100 GPUs. 
See the summary in Table~\ref{tab:hyperparameters}.
\begin{table*}[t]
\centering
\caption{Training hyperparameters of \textsc{AudioRubrics}.}
\label{tab:hyperparameters}
\resizebox{0.6\columnwidth}{!}{
\begin{tabular}{ll}
\toprule
\textbf{Hyperparameter} & \textbf{Value} \\
\midrule
Algorithm & GRPO \\
Base Model & Qwen2.5-Omni-7B \\
Max Prompt Length & 4096 \\
Max Response Length & 1024 \\
KL Coefficient ($\beta_{\mathrm{KL}}$) & 0.001 \\
Learning Rate & 1e-6 \\
LR Scheduler & Constant \\
Sampling Temperature & 1.0 \\
Batch Size & 8 \\
Rollout Number ($G$) & 8 \\
Training Steps & 400 \\
Precision & bf16 \\
\midrule
Accuracy / Format Reward Weight & 0.9 / 0.1 \\
Rubric Reward Weight ($\gamma$) & 0.5 \\
Overthinking Penalty Weight ($\delta$) & 0.15 \\
Target Reasoning Length ($L$) & 256 \\
Rubrics Kept per Group (top-$M$) & 5 \\
New Rubrics per Group & 3 \\
Rubric Generator / Judge & Gemini-3.1-Pro \\
\bottomrule
\end{tabular}
}

\end{table*}

\section{Qualitative Analysis}
\subsection{Case Study}
To understand how our reward design improves audio reasoning, we conduct a qualitative analysis in Figure~\ref{fig:case-study-uncle} where the model must infer a speaker's communicative intent. The GRPO baseline answers incorrectly, anchoring on the most lexically salient token in the audio and producing a one-line justification that restates this surface topic without reasoning about the underlying intent. In contrast, \ours reconstructs the interaction step by step, tracks the evolving dialogue state, and rules out each distractor against the conversational flow rather than against isolated keywords. We attribute this behavior to our evolving rubric-based training process reward, which supplies fine-grained credit to intermediate reasoning steps and thereby incentivizes the model to ground its answer in the full context rather than in a single salient cue. 
\begin{figure*}[h]\centering
\begin{mybox}[Case Study: GRPO and \ours]
{\small
\textbf{Question:} Why does the uncle say the last sentence in the audio?

\medskip
\textbf{Choices:}
\begin{itemize}\setlength\itemsep{1pt}
  \item[A.] Because he wants to explain that he is imitating Superman
  \item[B.] Because he thinks his nephew is praising him
  \item[C.] Because he feels his nephew doesn't understand his performance
  \item[D.] Because he thinks Superman is cool
\end{itemize}
\textbf{Ground Truth:} \textbf{B}

\medskip\hrule\medskip
\textbf{\textcolor{red!70!black}{GRPO (Baseline) — Predicted A \ding{55}}}\\
The uncle says the last sentence because he wants to explain that he is
imitating Superman.

\medskip
\textbf{\textcolor{green!50!black}{AudioRubrics (Ours) — Predicted B \ding{51}}}\\
On careful listening, the audio features a nephew asking to hear an
impression, then the uncle begins it (groaning ``No, not the kryptonite!'').
The nephew corrects him, noting it is Superman, to which the uncle responds
``Thank you, I've been practicing a lot.'' This indicates the uncle says the
last sentence because he \emph{mistakenly believes his nephew is praising}
his performance. Choices A and D do not match the conversational context,
and C is not supported by the dialogue flow.
}
\end{mybox}
\caption{Qualitative case study. The GRPO baseline latches onto the surface topic (Superman) and answers A,
whereas our model tracks the speaker's \emph{intent} across the dialogue and
correctly infers that the uncle misreads the correction as praise.}
\label{fig:case-study-uncle}
\end{figure*}
\subsection{Comparison between Initial and Evolved Rubrics}
\label{sec: case rubrics}
Figure~\ref{fig:rubric-evolution-disambig} compares the two rubric sources on a location example. The kept initial rubrics cover easy sub-tasks---grounding the sound and ruling out the water options---that both right and wrong rollouts already pass. The evolved rubrics instead come from the rollouts and target the hard split the answer turns on, \emph{field} vs.\ \emph{woods}, which the generator weights as highly as the grounding rubric. Thus the evolved rubrics add the question-specific discrimination that the fixed set lacks.
\begin{figure*}[h]\centering
\begin{mybox}[Case Study: Rubric Co-Evolution]
{\small
\textbf{Question:} Where does the audio take place?

\medskip
\textbf{Choices:}
\begin{itemize}\setlength\itemsep{1pt}
  \item[A.] aquatic
  \item[B.] at sea
  \item[C.] field
  \item[D.] woods
\end{itemize}
\textbf{Ground Truth:} \textbf{C}

\smallskip
\textit{(Footsteps crunch through dry vegetation in an open outdoor soundscape with faint birdsong---no water sounds.)}

\medskip\hrule\medskip
\textbf{\textcolor{blue!60!black}{Retained Static Rubrics}}\\
Two of the five initial criteria still discriminate and are kept, covering the \emph{easy} sub-tasks---grounding the evidence and ruling out the water options:
\begin{itemize}\setlength\itemsep{1pt}
  \item \textbf{[Auditory Evidence Grounding]} ($w{=}0.35$) identifies the footsteps crunching through dry vegetation as the primary evidence.
  \item \textbf{[Distractor Elimination]} ($w{=}0.15$) rules out ``aquatic''/``at sea'' by the complete absence of water-related sounds.
\end{itemize}
The other three generic static criteria---reasoning clarity, focus, and terminology---have saturated and are pruned by the variance filter.

\medskip
\textbf{\textcolor{green!50!black}{Newly Evolved Rubrics}}\\
The generator distills two criteria that target the discrimination the fixed rubrics never asked for, and gives the harder one the \emph{top} weight:
\begin{itemize}\setlength\itemsep{1pt}
  \item \textbf{[Woodland vs.\ Field Disambiguation]} ($w{=}0.35$) explicitly differentiates ``field'' from ``woods'' by citing specific acoustic differences, such as spatial acoustics (open vs.\ dense) or the nature of the rustling.
  \item \textbf{[Background Environmental Cues]} ($w{=}0.15$) identifies secondary sounds---bird chirping, wind noise---to further corroborate the outdoor terrestrial setting.
\end{itemize}

}
\end{mybox}
\caption{Qualitative case study of rubric co-evolution. Two static rubrics are retained for the easy sub-tasks, while the generator adds the fine-grained ``field vs.\ woods'' disambiguation---the crux of the question---and weights it as highly as the retained grounding criterion.}
\label{fig:rubric-evolution-disambig}
\end{figure*}

\section{Additional Experimental Results}
\subsection{Analysis on Response Length}
\label{sec:training_dynamics}
\begin{wrapfigure}{r}{0.4\textwidth}
	\centering
	\includegraphics[width=\linewidth]{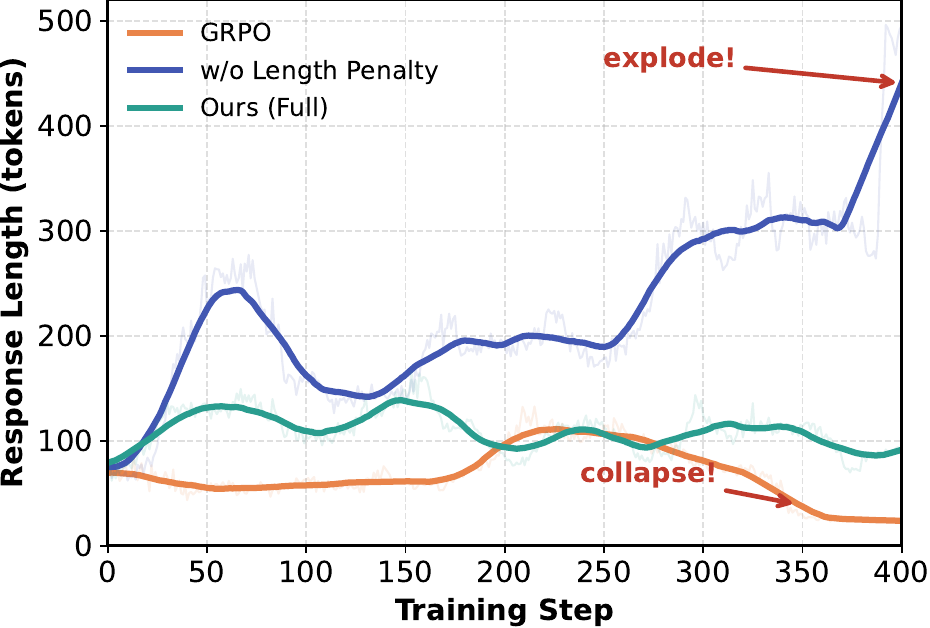}
	\caption{Average response length during training under three reward configurations. Vanilla GRPO collapses toward near-zero length, removing the length penalty leads to unbounded growth, and our full method maintains a stable length throughout.}
	\label{fig:length_dynamics}
\end{wrapfigure}
Figure~\ref{fig:length_dynamics} tracks the average response length over the course of training under three reward configurations, revealing two opposite
pathologies that our full design avoids. Under vanilla GRPO (orange), the response length steadily collapses: after a brief initial plateau, it decays
toward near-zero by the end of training, indicating that the policy learns to
shortcut to terse, under-reasoned answers that maximize the outcome reward
without producing an intermediate reasoning trace. Removing the length penalty
from our objective (blue) induces the opposite failure mode: the response
length grows in an unbounded, increasingly unstable manner and ultimately
explodes, as the model is rewarded for ever-longer generations regardless of
their quality, inflating verbosity and compute cost without commensurate gains
in correctness. In contrast, our full method (green) maintains a stable response
length throughout training, fluctuating within a narrow band rather than
drifting toward either extreme. This stability indicates that the rubric-based
process reward and the length regularization act in complementary ways: the
former supplies dense credit for substantive reasoning steps, preventing the
degenerate collapse seen under GRPO, while the latter discourages uninformative
padding, keeping the generated rationales concise yet sufficiently detailed. The
result is a training process that converges to a well-calibrated reasoning
length, which we find correlates with the accuracy improvements reported in
Table~\ref{tab:mmsu_mmau}.

\subsection{Robustness to Different Model Scale}
We further apply our method to the 3B model to demonstrate its generalizability and robustness across model scales. We present the results in
Table~\ref{tab:ablation_3b}, comparing against the GRPO baseline trained under
the identical setting. From the table, we observe that our method consistently
outperforms GRPO across all three benchmarks, improving MMAU, MMAR, and MMSU by
$1.23\%$, $1.72\%$, and $3.65\%$ relative, respectively. The gains persist even at this smaller scale, where the reduced model capacity leaves less
room for reasoning-oriented rewards to take effect. The consistent improvement indicates that our rubric-based process reward provides a robust learning signal across different model scales.
\begin{table}[t]
\centering
\caption{Results on the 3B model. $\Delta_{rel}^{\%}$ denotes the relative
improvement over the GRPO baseline.}
\label{tab:ablation_3b}
\resizebox{0.5\columnwidth}{!}{%
\begin{tabular}{l|cc|cc|cc}
\toprule
\multirow{2}{*}{\textbf{Method}}
 & \multicolumn{2}{c|}{\cellcolor{gMMAU}\textbf{MMAU}}
 & \multicolumn{2}{c|}{\cellcolor{gMMAR}\textbf{MMAR}}
 & \multicolumn{2}{c}{\cellcolor{gMMSU}\textbf{MMSU}} \\
\cmidrule(lr){2-3}\cmidrule(lr){4-5}\cmidrule(lr){6-7}
 & \textbf{ACC} & $\Delta_{rel}^{\%}$
 & \textbf{ACC} & $\Delta_{rel}^{\%}$
 & \textbf{ACC} & $\Delta_{rel}^{\%}$ \\
\midrule
\rowcolor{barGray}\multicolumn{7}{c}{\textbf{3B}} \\
\midrule
GRPO  & 73.00 & --
      & 58.30 & --
      & 60.26 & -- \\
\ours & \textbf{73.90} & \cellcolor{green!15}$\uparrow$1.23
      & \textbf{59.30} & \cellcolor{green!22}$\uparrow$1.72
      & \textbf{62.46} & \cellcolor{green!41}$\uparrow$3.65 \\
\bottomrule
\end{tabular}
}
\end{table}

\section{Prompts}
This section lists the full prompts used throughout our pipeline: the inference prompt for benchmark evaluation, the static rubric generation prompt used for per-question initialization, the evolving-rubric system and judging prompts corresponding to the first $\Phi$ call of Algorithm~\ref{alg:rubric}, and the weight assignment prompt corresponding to the second call.

\begin{figure*}[h]\centering
\begin{mybox}[Inference Prompt (MCQ Evaluation)]
\begin{verbatim}
[System]
You are an expert audio understanding assistant. Listen carefully and answer
multiple-choice questions. Always think step by step inside <think> tags, then
give the final answer letter inside <answer> tags.

[User]  (audio attached as input_audio)
Listen to the audio carefully and answer the following multiple-choice question.

Question: {question}

Choices:
A. {choice_A}
B. {choice_B}
...

First, reason step by step inside <think> ... </think> tags.
Then output ONLY the letter (A, B, C, ...) of the correct answer inside
<answer> ... </answer> tags.

Example format:
<think>
Your reasoning here.
</think>
<answer>B</answer>
\end{verbatim}
\end{mybox}
\caption{Inference prompt used for all benchmark evaluation (greedy decoding, temperature $=0$).}
\label{fig:prompt-infer}
\end{figure*}

\begin{figure*}[h]\centering
\begin{mybox}[Static Rubric Generation Prompt]
\begin{verbatim}
You are an expert in audio understanding evaluation and rubric design.
Your task is to listen to a given audio clip and analyze the question-answer
pair, then generate exactly 5 evaluation rubrics that together assess response
quality for this question. The reference answer is only one possible response
from a student (not necessarily a good one).

# Input Data
[Audio]: <provided as audio attachment>
[Question]: {question}
[Reference Answer]: {response}

# The 5 rubric categories (use EACH category exactly once)
1. Auditory Evidence Identification & Grounding
2. Cross-Cue Verification & Distractor Elimination
3. Reasoning Clarity & Flow
4. Reasoning Focus & Efficiency
5. Domain-Specific Audio Techniques

# Rules
- Exactly 5 rubrics, ONE per category; no category repeats.
- Each rubric is binary (Yes/No) and MUST reference CONCRETE audio content
  actually heard in THIS clip or a specifically named distractor option.
- Each rubric carries a weight in (0,1); the 5 weights MUST sum to 1.0.
- DO NOT evaluate final-answer correctness (handled by a separate
  ground-truth reward).

# Output (JSON)
{
  "question_domain": "audio_reasoning/<sub-domain>",
  "rubrics": [{"category": ..., "criterion": ..., "weight": 0.xx} x5],
  "reference_score": x   # weights the reference would satisfy; must be < 0.5
}
\end{verbatim}
\end{mybox}
\caption{Prompt for generating the initial (static) per-question rubrics.}
\label{fig:prompt-static-rubric}
\end{figure*}

\begin{figure*}[h]\centering
\begin{mybox}[Evolving Rubric System Prompt]
\begin{verbatim}
You are an expert evaluator generating adaptive rubrics to assess model responses 
on audio tasks, which take audio + text as input and must reason over what they hear.

## Task
Identify the most discriminative criteria that distinguish high-quality from
low-quality audio reasoning responses. Capture subtle quality differences that
existing rubrics miss particularly those unique to grounding answers in actual 
acoustic evidence rather than text-only priors.

## Output Components
- Description: detailed, specific description of what makes a response excellent
  or problematic when reasoning over audio
- Title: concise abstract label (general, transferable across audio tasks)

## Categories
1. Positive Rubrics: excellence indicators distinguishing superior reasoning
2. Negative Rubrics: critical flaws definitively degrading reasoning quality

## Audio-Reasoning Quality Dimensions
- Acoustic Grounding: evidence actually present (words, sounds, prosody,
  timestamps) vs. plausible guesses; resistance to hallucinating audio content;
  faithfulness to what is audible vs. text/linguistic priors.
- Temporal & Sequential Reasoning: event ordering, durations/timestamps,
  tracking state changes across the timeline.
- Multi-Source Disentanglement: overlapping speakers, fore/background,
  speaker attribution/diarization, simultaneous events.
- Paralinguistic & Non-Lexical Reasoning: emotion, sarcasm, urgency, hesitation
  from prosody; what is said vs. how it is said; laughter/sighs/silence.
- Non-Speech Acoustic Understanding: sound-event ID, acoustic scene, music
  analysis (genre, tempo, key, instrumentation).
- Counting & Quantification: speakers, repetitions, distinct events, beats.
- Causal & Inferential Reasoning Over Audio: cues -> plausible causes; avoid
  spurious inference from a single ambiguous cue.
- Cross-Modal Integration: use the question to focus on relevant audio; do not
  let text priors override audio evidence when they conflict.

## Core Guidelines
1. Discriminative Power: only criteria that meaningfully separate the ACTUAL
   responses provided; exclude generic criteria ("is helpful").
2. Novelty & Non-Redundancy: never duplicate existing rubrics; add granular
   criteria if existing are broad; return empty lists if already comprehensive.
3. Avoid Mirror Rubrics: never create positive/negative versions of the same
   criterion; choose only the more discriminative direction.
4. Conservative Negative Rubrics: clear, observable failure modes (not mere
   absence of excellence); audio hallucination is a key negative axis.

## Selection Strategy
- Quantity: 1-5 total rubrics (fewer high-quality > many generic).
- More positive when responses are grounded but unsophisticated; more negative
  when systematic failures are present; empty when existing rubrics are comprehensive.

## Output Format (JSON)
{
  "question": "<original question verbatim>",
  "positive_rubrics": [{"description": ..., "title": ...}],
  "negative_rubrics": [{"description": ..., "title": ...}]
}

## Critical Reminders
- Each rubric must distinguish between the actual provided responses.
- Focus on objective, audio-grounded criteria a verifier could check against the audio.
- Treat audio-content hallucination and text-only reasoning that ignores the
  audio as first-class failure modes. 
\end{verbatim}
\end{mybox}
\vspace{-10pt}
\caption{Full system prompt that drives per-step rubric evolution (Call 1).}
\label{fig:prompt-evolve-sys}
\end{figure*}

\begin{figure*}[h]\centering
\begin{mybox}[Evolving Rubric Generation Judging Prompt]
\begin{verbatim}
## Question
{question}

## Ground-truth final answer
{gt_letter}

## Existing rubrics
S1 ({category}): {criterion}
...
S5 ({category}): {criterion}

## Candidate Responses (N rollouts of the same prompt)
--- T1  [final answer: {ans} -> CORRECT / WRONG (chose X)] ---
{reasoning trace extracted from <think>...</think>}
--- TN ... ---

## Per-rollout correctness summary
T1: CORRECT ... TN: WRONG (chose B)
Note: a rollout's final-answer correctness is context only. Generate rubrics
that distinguish *reasoning quality*, not just correctness -- a rollout may be
CORRECT by guessing yet deserve a low score; a rollout may be WRONG yet show
good acoustic grounding.

## Required Output (strict JSON, no code fences)
Generate AT MOST {max_new} NEW rubrics non-redundant with S1..S5. Each new
rubric has a 'polarity': 'positive' (satisfied = good) or 'negative'
(satisfied = bad/flaw present). Then judge ALL rubrics (static + new) against
EACH of the 8 responses T1..T8 using exactly 'Yes' or 'No'.

Schema:
{"new_rubrics": [{"id":"N1","title":"<label>","description":"<criterion>",
                  "polarity":"positive|negative"}, ...],
 "judgments": {"S1":["Yes"|"No", x8], ..., "S5":[...], "N1":[...], ...}}
\end{verbatim}
\end{mybox}
\caption{Call 1: jointly generate new candidate rubrics and judge all rubrics against the 8 rollouts.}
\label{fig:prompt-evolve-call1}
\end{figure*}

\begin{figure*}[h]\centering
\begin{mybox}[Rubric Weight Assignment Prompt]
\begin{verbatim}
You are an expert audio-reasoning evaluator. You will receive an audio clip, a
question, and a list of K binary rubrics that have been chosen as the most
discriminative for assessing model responses on this clip. Listen to the audio,
read the question, and decide how the K rubrics should be weighted: rubrics that
more directly probe whether the response is grounded in this specific audio
should receive higher weight. Output exactly K positive numbers that sum to 1.0.
Strict JSON output: {"weights": [w1, w2, ..., wK]}

## Question
{question}

## Kept Rubrics
R1: {rubric_text}
R2: {rubric_text}
...
RK: {rubric_text}

## Required Output
Return ONLY: {"weights": [..K floats summing to 1.0..]}
\end{verbatim}
\end{mybox}
\caption{Call 2: assign normalized importance weights to the top-$K$ kept rubrics.}
\label{fig:prompt-evolve-call2}
\end{figure*}

\begin{figure*}[h]\centering
\begin{mybox}[Rubric Judging Prompt]
\begin{verbatim}
You are an expert audio response evaluator. Listen to the attached audio, read
the question, candidate response, and 5 binary rubrics. For each rubric, decide
whether the candidate response satisfies it (Yes) or not (No).

[Question]: {question}

[Candidate Response]:
{response}

[Rubrics to evaluate]:
R1 ({category}, weight {w}): {criterion}
...
R5 ({category}, weight {w}): {criterion}

Return strictly JSON (no code fences):
{"R1": "Yes|No", "R2": "Yes|No", "R3": "Yes|No", "R4": "Yes|No", "R5": "Yes|No"}
\end{verbatim}
\end{mybox}
\caption{Prompt for judging whether a response satisfies each rubric.}
\label{fig:prompt-judge}
\end{figure*}

\begin{algorithm}[t]
\caption{Evolving rubric reward for one prompt group}
\label{alg:rubric}
\begin{algorithmic}[1]
\Require audio $A$, question $Q$, ground-truth answer $y^\star$, rollouts $\{o_i\}_{i=1}^{G}\sim\pi_\theta$, retained rubric set $\mathcal{R}_{\text{prev}}$ from the previous iteration (initialized to the weighted initial rubrics $\mathcal{R}_0=\{(r_k,w_k)\}_{k=1}^{K}$ from Eq.~\ref{eq:init} at the first iteration), rubric budget $N_{\text{new}}$, keep size $M$, generator--judge $\Phi$
\Ensure rubric rewards $\{R^{\text{rub}}_i\}_{i=1}^{G}$; updated rubric set $\mathcal{K}$ carried to the next iteration
\Statex \textit{(i) Elicitation and judging}
\State $\big(\mathcal{R}_{\text{new}},\{j_{k,i}\}\big)\gets\Phi\big(A,Q,y^\star,\{o_i\}_{i=1}^{G},\mathcal{R}_{\text{prev}};\,N_{\text{new}}\big)$ \Comment{$\le N_{\text{new}}$ polarity-tagged new rubrics; verdicts $j_{k,i}\!\in\!\{0,1\}$ for all rubrics}
\State $\mathcal{R}\gets\mathcal{R}_{\text{prev}}\cup\mathcal{R}_{\text{new}}$
\For{each rubric $r_k\in\mathcal{R}$ and each rollout $i=1,\dots,G$}
  \State $b_{k,i}\gets j_{k,i}$ \textbf{if} $r_k$ is positive \textbf{else} $1-j_{k,i}$ \Comment{polarity normalization, Eq.~\ref{eq:polarity}}
\EndFor
\Statex \textit{(ii) Variance filtering}
\State $s_k\gets\operatorname{std}_i\{b_{k,i}\}_{i=1}^{G}$ for each $r_k\in\mathcal{R}$
\State $\mathcal{K}\gets\{\,r_k\in\mathcal{R}:\,s_k>0\,\}$ \Comment{drop criteria that all rollouts pass or all fail}
\If{$\mathcal{K}=\emptyset$}
  \State \Return $R^{\text{rub}}_i\gets 0.5$ for all $i$ \Comment{neutral fallback: a constant reward contributes no advantage}
\EndIf
\State $\mathcal{K}\gets$ top-$M$ rubrics of $\mathcal{K}$ ranked by $s_k$ \Comment{keep the most discriminative criteria; all if fewer than $M$ survive; $\mathcal{K}$ becomes $\mathcal{R}_{\text{prev}}$ next iteration}
\Statex \textit{(iii) Weighting and scoring}
\State $\{w_k\}_{k\in\mathcal{K}}\gets\Phi\big(A,Q,\mathcal{K}\big)$ \Comment{positive, $\sum_{k\in\mathcal{K}}w_k=1$; supersedes the initial weights}
\State \Return $R^{\text{rub}}_i\gets\sum_{k\in\mathcal{K}}w_k\,b_{k,i}$ for $i=1,\dots,G$ \Comment{Eq.~\ref{eq:rubric-reward}; enters $R_i$ in Eq.~\ref{eq:reward}}
\end{algorithmic}
\end{algorithm}

\end{document}